\documentclass[longauth]{aa} 

\usepackage{graphicx}
\usepackage{txfonts}
\usepackage{hyperref}
\usepackage{tikz}
\usepackage{upgreek}

\begin{document} 

   \title{Characterizing the radio emission from the binary galaxy cluster merger Abell 2146}


   \author{D. N. Hoang\inst{1}, T. W. Shimwell\inst{2,1}, R. J. van Weeren\inst{1}, H. J. A. R\"{o}ttgering\inst{1}, A. Botteon\inst{3,4}, G. Brunetti\inst{3}, \mbox{M. Br\"uggen\inst{5},} R. Cassano\inst{3}, J. Hlavacek-Larrondo\inst{6}, M.-L. Gendron-Marsolais\inst{6}, and A. Stroe\inst{7}
          }

   \institute{Leiden Observatory, Leiden University, PO Box 9513, NL-2300 RA Leiden, the Netherlands\\
              \email{hoang@strw.leidenuniv.nl}
         \and
             Netherlands Institute for Radio Astronomy (ASTRON), P.O. Box 2, 7990 AA Dwingeloo, The Netherlands 
                \and
                        INAF- Istituto di Radioastronomia, via P. Gobetti 101, 40129, Bologna, Italy 
                \and
                        Dipartimento di Fisica e Astronomia, Universit\`{a} di Bologna, via P. Gobetti 93/2, I-40129 Bologna, Italy
                \and
                        Hamburger Sternwarte, University of Hamburg, Gojenbergsweg 112, 21029 Hamburg, Germany
                \and
                        D\'{e}partement de Physique, Universit\'{e} de Montr\'{e}al, Montr\'{e}al, QC H3C 3J7, Canada
                \and
                        European Southern Observatory, Karl-Schwarzschild-Str. 2, D-85748 Garching, Germany
             }
   \titlerunning{Characterizing the radio emission from Abell 2146}
   \authorrunning{D. N. Hoang, et. al.}
   \date{Received... 2018; ... 2018}
        
 
  \abstract
   {Collisions of galaxy clusters generate shocks and turbulence in the intra-cluster medium (ICM). The presence of relativistic particles and magnetic fields is inferred through the detection of extended synchrotron radio sources such as haloes and relics and implies that merger shocks and turbulence are capable of \mbox{(re-)accelerating} particles to relativistic energies. However, the precise relationship between merger shocks, turbulence, and extended radio emission is still unclear. Studies of the most simple binary cluster mergers are important to help understand the particle acceleration in the ICM.}
   {Our main aim is to study the properties of the extended radio emission and particle acceleration mechanism(s) associated with the generation of relativistic particles in the ICM.}
   {We measure the low-frequency radio emission from the merging galaxy cluster Abell 2146 with LOFAR at 144 MHz. We characterize the spectral properties of the radio emission by combining these data with data from archival Giant Metrewave Radio Telescope (GMRT) at 238 MHz and 612 MHz and Very Large Array (VLA) at 1.5 GHz.}
   {We observe extended radio emission at 144 MHz behind the NW and SE shocks. Across the NW extended source, the spectral index steepens from $-1.06\pm0.06$ to $-1.29\pm0.09$ in the direction of the cluster centre. This spectral behaviour suggests that a relic is associated with the NW upstream shock. The precise nature of the SE extended emission is unclear. It may be a radio halo bounded by a shock or a superposition of a relic and halo. At 144 MHz, we detect a faint emission that was not seen with high-frequency observations, implying a steep ($\alpha<-1.3$) spectrum nature of the bridge emission. 
        }
   {Our results imply that the extended radio emission in Abell 2146 is probably associated with shocks and turbulence during cluster merger. The relativistic electrons in the NW and SE may originate from fossil plasma and thermal electrons, respectively. }
   \maketitle
%
\section{Introduction}
\label{sec:intro}

Extended radio synchrotron emission ($\sim\text{Mpc}$-scale) in clusters is generally associated with large-scale shocks and turbulence that are generated in the intra-cluster medium (ICM) during the formation of galaxy clusters (see e.g. \citealt{Ferrari2008,Luigina2012,Brunetti2014} for reviews). The detection of this emission reveals the presence of relativistic particles (i.e. cosmic rays; CRs) and magnetic fields in the ICM. Based on its physical properties, the extended emission is commonly classified as either a relic or a halo \citep[e.g.][]{Kempner2004}. Radio relics are elongated, highly polarised, steep spectrum sources that are usually observed at the cluster periphery.  Radio haloes are spherically shaped, apparently unpolarised sources that are found in the central regions of clusters. Owing to the short cooling timescale of the radio-emitting relativistic electrons through synchrotron and inverse-Compton (IC) energy losses in $\sim\upmu\text{G}$ magnetic fields, the $\sim\text{Mpc}$ scale of radio relics and haloes implies that CRs must be \mbox{(re-)accelerated} in situ \citep{Jaffe1977}.

The favoured mechanism of particle acceleration in haloes is where CRs are energised by turbulence during cluster mergers  \citep[e.g.][]{Brunetti2001,Petrosian2001a,Fujita2003,Cassano2005,Brunetti2007a,Brunetti2016,Pinzke2017}. The collisions of the relativistic protons and thermal protons in the ICM (i.e. the hadronic model), which produce secondary relativistic electrons, may also contribute to the observed radio emission  \citep[e.g.][]{Dennison1980a,Blasi1999,Dolag2000,Miniati2001,Pfrommer2004a,Pfrommer2008,Keshet2010,EnBlin2011}. However, pure hadronic models for radio haloes are severely challenged by limits from Fermi-LAT observations  \citep[e.g.][]{Jeltema2011,Brunetti2012,Zandanel2014a,Ackermann2010,Ackermann2016b}, although hadronic models in which secondary particles are \mbox{re-accelerated} by turbulence might be still possible to explain the radio emission in haloes \cite{Brunetti2011,Brunetti2017,Pinzke2017}. The generation and dissipation of turbulence in the ICM is a complicated process and involves a complex chain of mechanisms \citep[see e.g.][for review]{Brunetti2014}. For example, the gravitational potential induced by the motion of dark matter sub-haloes can cause mixing of the gas and stretching of the magnetic fields in the ICM, which generates instabilities at different scales \citep[e.g.][]{Cassano2005,Fujita2004,ZuHone2013,Miniati2014}. Also merger shocks driven in the ICM can generate turbulence at different scales via baroclinic and compressive processes\citep[e.g.][]{Iapichino2008b,Iapichino2008a,Iapichino2011,Vazza2017}.

The relativistic electrons in relics are thought to be more locally generated through Fermi-I acceleration by cluster-scale shocks during the collisions of sub-clusters/groups \citep[e.g.][]{Ensslin1998,Roettiger1999a}. The observation evidence for this is that radio relics have been observed at the location of some X-ray shock fronts \citep[e.g.][]{Shimwell2015,Botteon2016a,Eckert2016,Akamatsu2017,VanWeeren2016b,VanWeeren2017,Urdampilleta2018}. The steepening of the radio spectral index across the width of the elongated relics provides additional evidence that relativistic electrons in relics are \mbox{(re-)accelerated} at shock fronts and lose their energy in the post-shock region due to synchrotron and IC energy losses \citep[e.g.][]{Giacintucci2008,VanWeeren2010a,Stroe2013a,Hoang2017a,Hindson2014}. Moreover, the alignment of magnetic field vectors along the length of some relics implies a compression of the magnetic fields in these regions, which is an expected consequence of a passing shock front  \citep[e.g.][]{Bonafede2009,Bonafede2012,Kale2012b,deGrasperin2014,Pearce2017,Hoang2018a}.

Despite the observational evidence associating radio relics with shocks, radio relics have not been firmly detected at all known  strong ($\mathcal{M}\approx2-3$) cluster shocks. For example, relics are not obviously associated with shocks in Abell 520 \citep[e.g.][submitted]{Markevitch2005,Govoni2001c,Vacca2014,Hoang2018a},  the western edge of the Bullet cluster \citep[$1\text{E}\,065-558$; e.g.][]{Markevitch2002a,Shimwell2014}, Abell 665 \citep[e.g.][]{Feretti2004,Dasadia2016}, and Abell 2146   \citep[e.g.][]{Russell2010,Hlavacek-Larrondo2017}. 
In these regions, extended radio emission is either undetected at the shocks (e.g. Abell 665), the spectrum of the observed radio emission does not steepen in the post-shock region (e.g. the Bullet cluster), or the distribution of spectral index has not been obtained (e.g. Abell 2146). Moreover, polarimetric measurements have not been reported for most of these clusters, except for the Bullet cluster. The non-detection of radio relics at these shocks would challenge the shock-related formation model of radio relics, however observational limitations should be carefully sorted out. 
%
To address this problem, we performed new low-frequency radio observations of galaxy clusters with prominent shocks but debated or no radio relic emission, i.e. Abell 520 \citep[][submitted]{Hoang2018a}, Abell 665, and Abell 2146 (this paper).

\section{Galaxy cluster Abell 2146}
\label{sec:a2146}

Abell 2146 (hereafter A2146; $z=0.232$) is a binary merging galaxy cluster in which the first core passage occurred approximately 0.24-0.28 Gyr ago in the plane of the sky. \citep[e.g.][]{Russell2010,Russell2011,Russell2012,Rodriguez-Gonzalvez2011,Canning2012,White2015,King2016,Coleman2017,Hlavacek-Larrondo2017}. The total mass of A2146 is estimated to be $M_\text{500}=(4.04\pm0.27)\times10^{14}\,\text{M}_\odot$ \citep{Planck2015}. \textit{Chandra} X-ray observations revealed a NW-SE elongation and a highly disturbed morphology of the thermal ICM \citep[][]{Russell2010}. Detailed analysis of the X-ray surface brightness (SB) and temperature distribution revealed a bow shock in the SE region and an upstream shock in the NW region \citep{Russell2010,Russell2012}. 
From the jumps in the X-ray derived density profiles, \cite{Russell2012} estimated Mach numbers of $\mathcal{M}_\text{SE}^\text{X}=2.3\pm0.2$ and $\mathcal{M}_\text{NW}^\text{X}=1.6\pm0.2$ for the SE and NW shocks, respectively. Even though A2146 is a highly disturbed cluster with clear shock fronts, initially radio observations did not detect extended radio emission from the cluster \citep{Russell2011}. However, recently \cite{Hlavacek-Larrondo2017} discovered radio emission extending up to $\sim850\,\text{kpc}$ with deep VLA L-band observations. The $\sim30\arcsec$ resolution VLA L-band image shows two separated patches of extended emission in the NW and SE regions. Based on the location and morphology, \cite{Hlavacek-Larrondo2017} have suspected that the NW emission is a radio relic associated with the NW upstream shock. Although \cite{Hlavacek-Larrondo2017} have suggested that the SE emission might be a radio halo, they also speculated whether or not it may consist of a halo and a relic that could not be separated with their low-resolution VLA image. However, for both the NW and SE regions key observational evidence to connect the extended radio emission (i.e. relics) with the shocks (i.e. spectral steepening in post-shock regions and alignment of magnetic fields at the shock fronts) is still missing. Thus, the precise nature of the diffuse radio emission in A2146 still needs to be determined. 

%
%
%

In this paper, we present deep LOFAR $120-168$ MHz observations of A2146. We aim to map the extended emission from A2146 at 144 MHz with a high resolution ($\lesssim15\arcsec$). To study spectral properties of the radio emission from the cluster, we combine the LOFAR 144 MHz data with the existing VLA $1-2$ GHz data presented in \cite{Hlavacek-Larrondo2017} and the archival GMRT $222-254$ MHz and $596-628$ MHz data. We assume the cosmological parameters, $H_{0}=70$ km s$^{-1}$  Mpc$^{-1}$, $\Omega_{M}=0.3$, and $\Omega_{\Lambda}=0.7$. In this cosmology, $1\arcmin$ corresponds to $221.88$ kpc at $z=0.232$.

\section{Observations and data reduction}
\label{sec:obs_red}

\subsection{LOFAR 144 MHz}
\label{sec:obs_red_lofar}

A2146 was observed for a total of 16 hours split over two dates (ObsID: L589831 and L6311955) with LOFAR (\citealt{VanHaarlem2013}). One of the observations (i.e. L589831) forms part of the LOFAR Two-metre Sky Survey (LoTSS; \citealt{Shimwell2017}) and has a pointing centre of $1.3^\circ$ from the target. The other is a targeted observation centred on A2146. A summary of the observations is given in Table \ref{tab:obs_para}. 

\begin{table*}
        \centering
        \caption{Observation details}
        \begin{tabular}{lcccc}
                \hline\hline
                Telescope                 &    LOFAR 144 MHz    & GMRT 238 MHz  &   GMRT 612 MHz   &      VLA 1.5 GHz      \\ \hline
                Project codes             & LC7\_024, DDT9\_001 &    20\_065    & 20\_065, 12DT009 &        12A-029        \\
                Configurations            &         $-$         &      $-$      &       $-$        &        B, C, D        \\
                Observing dates           &    Apr. 25, 2017;    & Jun. 05, 2011 &  Jun. 05, 2011;  &   Jun. 9, 2012 (B);   \\
                                          &    Dec. 7, 2017     &               &  Mar. 17, 2012   &  Mar. 16, 2012 (C);   \\
                                          &                     &               &                  &  Apr. 22, 2012 (C);   \\
                                          &                     &               &                  &   Jan. 27, 2013 (D)   \\
                Obs. IDs                  &  L589831, L631955   &     5369      &    5369, 5888    &          $-$          \\
                Calibrators               &   3C 295, 3C 196    &     3C 48      &      3C 286       &  3C 286, J1634+6245   \\
                Bandwidth (MHz)           &         48          &      32       &        32        &         1000          \\
                Total on-source time (hr) &         16          &      8.7      &        15        &         11.1          \\
                Integration time (s)      &          1          &      16       &        16        &    3 (B), 5 (C, D)    \\
                Correlations              &   XX, XY, YX, YY    &    RR, LL     &      RR, LL      & RR, RL, LR, LL (B, D) \\
                                          &                     &               &                  &      RR, LL (C)       \\
                Number of antennas        &         62          &      28       &      28, 30      &        15, 16         \\ \hline\hline
        \end{tabular}\\
        \label{tab:obs_para}
\end{table*}

We calibrated the LOFAR data to correct for direction-independent and direction-dependent effects using the facet calibration technique described in \cite{VanWeeren2016a} and \cite{Williams2016a} and summarised in this work for completeness. The direction-independent calibration (see  \cite{deGasperin2018a} for an overview of direction independent effects) is carried out using the $\mathtt{PreFactor}$\footnote{\url{https://github.com/lofar-astron/prefactor}} pipeline, which includes flagging radio frequency interference (RFI), removing contamination from bright sources in the distant side lobes (i.e. Cassiopeia A and Cygnus A), correcting the amplitude gain, the initial clock offsets, and XX-YY phase offsets. The calibration parameters were derived from 10 min observations of the primary calibrators 3C 196 and 3C 295 (Obs. IDs: L589831 and L6311955). The 3C 196 and 3C 295 models used to calibrate these data have integrated flux densities that are consistent with the \cite{Scaife2012} flux scale. After the target data are calibrated with the solutions derived from the calibrator observations, the data are phase calibrated against a wide-field sky model obtained from the TIFR GMRT 150 MHz All-sky Radio Survey: First Alternative Data Release \citep[TGSS-ADR1;][]{Intema2017}. The direction-dependent calibration is performed separately on each data set using the $\mathtt{Factor}$\footnote{\url{https://github.com/lofar-astron/factor}} pipeline, which aims to correct for ionospheric distortions and errors in the primary beam model to allow for accurate calibration in the direction of A2146.

To obtain continuum images, the data calibrated in the direction of A2146 were deconvolved with $\mathtt{MS-MFS}$ (multi-scale--multi-frequency) with $\mathtt{nterms=2}$ and $\mathtt{W-}$projection options in the Common Astronomy Software Applications package ($\mathtt{CASA}$) to account for the frequency dependence of the sky and non-coplanar effects \citep{Cornwell2005,Cornwell2008,Rau2011}. To map emission at different scales, we made continuum images with various weightings of the visibilities. All continuum images that were obtained from the observations were corrected for primary beam attenuations. To minimise the uncertainty in the LOFAR flux scale, the LOFAR images obtained from the observations L589831 and L6311955 are multiplied by factors of $1.02$ and $1.18$, respectively; where these scaling factors were obtained by comparing the integrated flux densities of nearby compact sources in the LOFAR images with those from the TGSS-ADR1 \citep{Intema2017}. The pixel values for the final combined continuum image are calculated as the average of the primary beam, corrected images, weighted by the local noise in each image. 

\subsection{GMRT 238 and 612 MHz}
\label{sec:red_gmrt}

The GMRT 238 and 612 MHz observations of A2146 were carried out on June 5, 2011 and March 17, 2012 (Obs. IDs: 5369 and 5888). Table \ref{tab:obs_para} provides more details on the observations. Each data set was separately calibrated using the Source Peeling and Atmospheric Modeling package ($\mathtt{SPAM}$; \citealt{Intema2009a,Intema2017}). This calibration mainly aims to correct for the ionospheric distortion, which includes an  interferometric phase delay. The amplitude gains were calibrated according to the flux scale in \cite{Scaife2012}. The flux scale error of 10\% is used for the GMRT observations \citep[e.g.][]{Chandra2004}. The final continuum images of A2146 at 612 MHz were obtained by averaging the primary beam corrected images from the different observations.

\subsection{VLA 1.5 GHz}
\label{sec:obs_red_vla}

A2146 was observed for a total of 11.1 hours with the VLA at 1.5 GHz in B, C, and D configurations on multiple dates in 2012 and 2013. The frequency bandwidth is divided into two intermediate frequency (IF) pairs, each of which has eight sub-bands of 64 MHz. The observations were carried out with full-polarisation settings. The integration time was set at three seconds for the B-configuration and five seconds for the C- and D-configurations. More details on the VLA observations are given in Table \ref{tab:obs_para}.

The calibration for each configuration of the VLA data sets was performed separately in $\mathtt{CASA}$ by \cite{Hlavacek-Larrondo2017}. For a full description of the procedure, we refer to \cite{Hlavacek-Larrondo2017}. 
To obtain continuum images, the calibrated data of the B-, C-, and D-configurations were combined and imaged with the $\mathtt{MS-MFS}$ and $\mathtt{W-}$projection algorithms in $\mathtt{CASA}$. 

\subsection{Spectral measurements}
\label{sec:red_spx}

We make a spectral index map to study the spectral energy distribution of the extended emission from A2146. The spectral index map is made with the LOFAR 144 MHz and VLA 1.5 GHz data. The LOFAR and VLA continuum images used for the spectral index mapping are imaged with the same parameters (i.e.  in the \textit{uv} range $0.12-65\,\text{k}\lambda$, Briggs $\mathtt{robust}$ weighting of 0.0, $\mathtt{outertaper}$ of $15\arcsec$ and $30\arcsec$ for the LOFAR and VLA, respectively, to obtain an approximate resolution of $30\arcsec$). The spectral index\footnote{In this paper, we use the convention $S\propto\nu^\alpha$} and the corresponding error for each pixel are estimated as
\begin{equation}
        \alpha = \frac{\ln\frac{S_1}{S_2}}{\ln\frac{\nu_1}{\nu_2}}  \qquad \text{and} \qquad \Delta\alpha=\frac{1}{\ln{\frac{\nu_1}{\nu_2}}}\sqrt{\left(\frac{\Delta S_1}{S_1}\right)^2+\left(\frac{\Delta S_2}{S_2}\right)^2},      
\end{equation}
where $\Delta S_i=\sqrt{(\sigma_i)^2+(f_\text{err}S_i)^2}$ with $i=[1,2]$ is the total error associating with the flux density measurement $S_i$; and the subscripts 1 and 2 stand for 144 MHz and 1.5 GHz, respectively. The total error is propagated from the flux scale uncertainty (i.e. $f_\text{err}=15\%$ and 5\% are commonly used for the LOFAR and VLA observations, respectively) and the image noise ($\sigma$).

\section{Results and discussion}
\label{sec:res_dis}

\begin{figure*}
        \centering
        \includegraphics[width=0.49\textwidth]{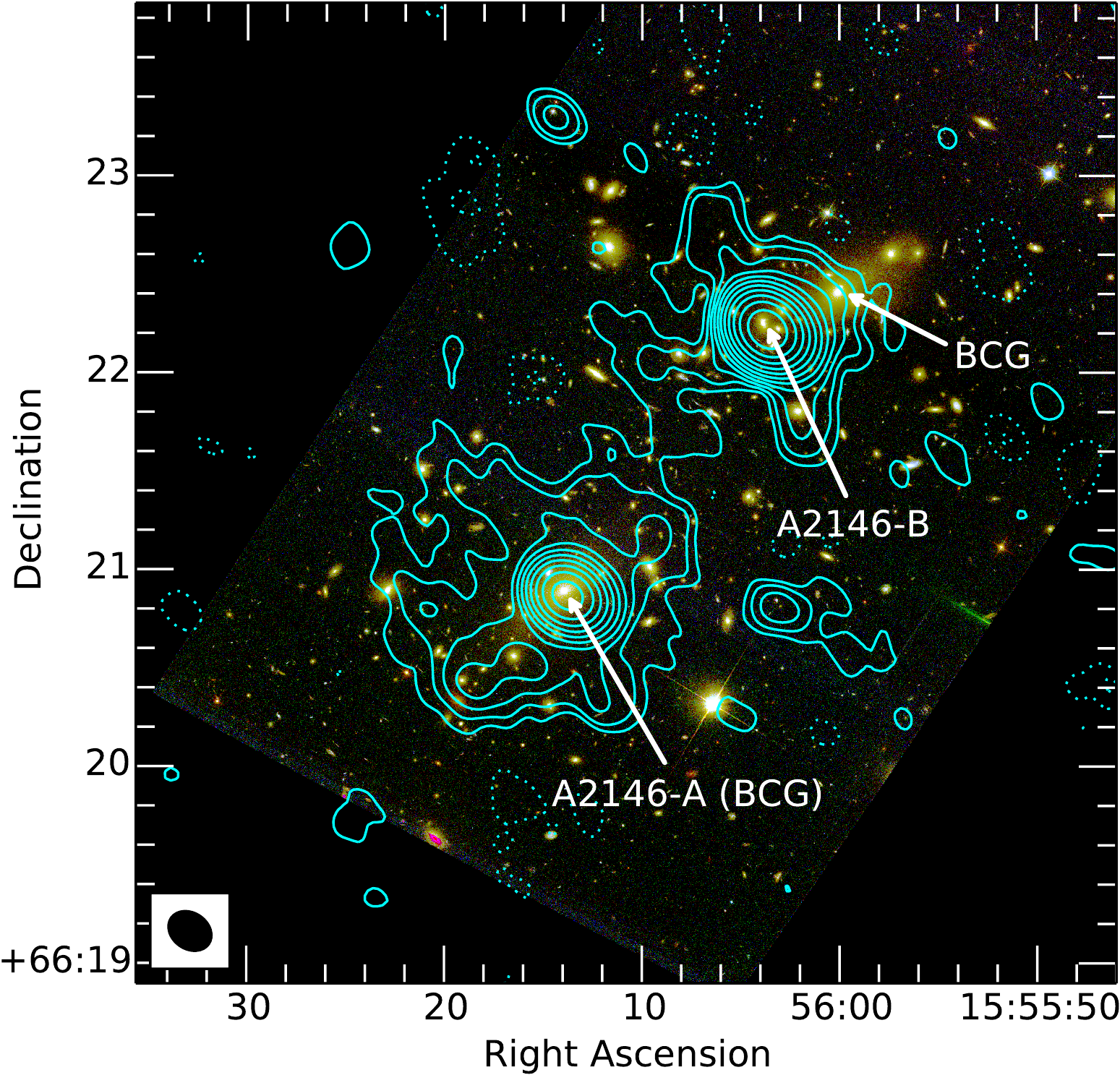} \hfill
        \includegraphics[width=0.49\textwidth]{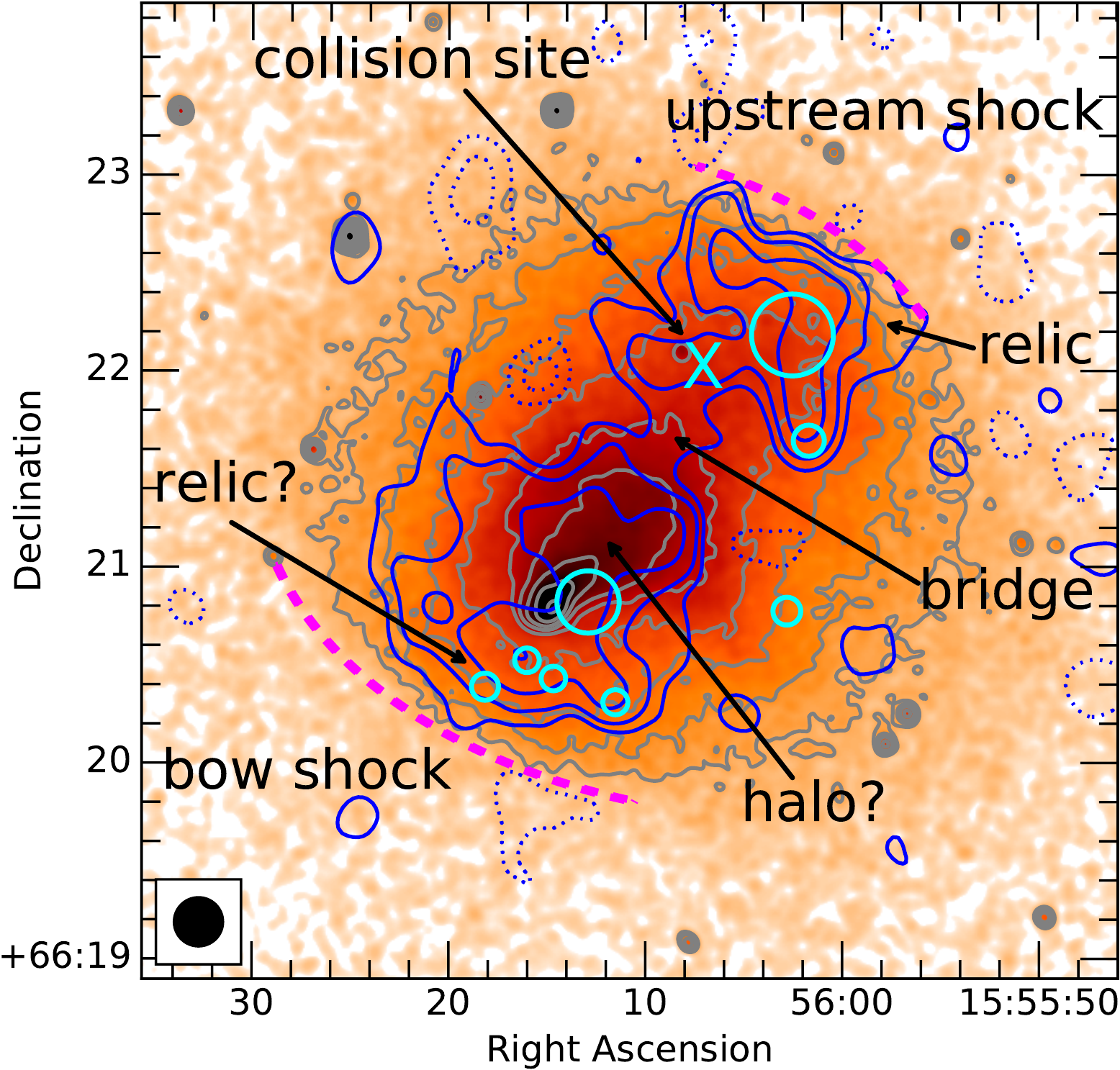}
        \caption{HST composite optical (left) and \textit{Chandra} (right) images overlaid with the LOFAR 144 MHz contours (blue in the right panel). The resolutions of the LOFAR contours shown in the bottom left corner are $14\arcsec\times11\arcsec$ ($P.A.=55^\circ$, left) and $15\arcsec\times15\arcsec$ (right). In the right panel, the compact radio sources indicated with cyan circles are subtracted in the \textit{uv} data. The magenta dashed lines show the shock locations. In both images, the LOFAR contours start from $\pm2.5\sigma$ (dotted negative), where $\sigma=135$ and  $160\,\upmu\text{Jy\,beam}^{-1}$ in the left and right panels, respectively. The \textit{Chandra} first contour is $2\times10^{-9}\,\text{cts\,cm}^{-2}\text{s}^{-1}\text{arcsec}^{-2}$. The next contours are spaced by a factor of $\sqrt{3}$.}
 \label{fig:hlres}
\end{figure*}

In Fig. \ref{fig:hlres}, we present LOFAR 144 MHz continuum images of A2146. In this section, we estimate flux densities and spectral index for the compact and extended sources in the cluster and discuss the implications.

\subsection{Radio galaxies}
\label{sec:res_dis_bcgs}

\begin{figure}
        \centering
        \includegraphics[width=0.49\textwidth]{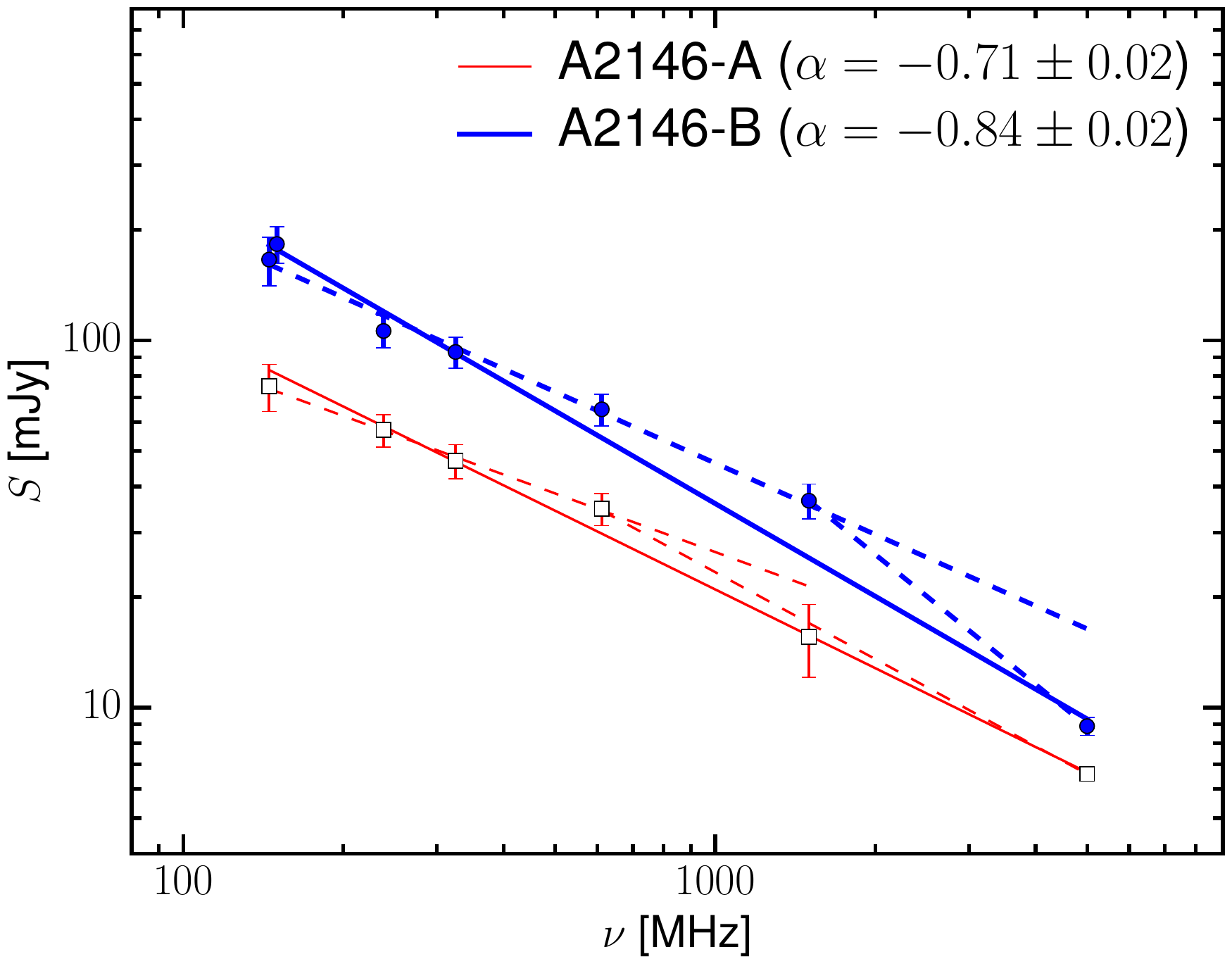}
        \caption{Spectral energy distribution between 144 MHz and 5 GHz for A2146-A and A2146-B. The flux densities are summarised in Table \ref{tab:int_flux}.}
        \label{fig:int_ps}
\end{figure}

A2146 is known to host two bright radio galaxies (namely A2146-A and A2146-B) that are located in the central regions of the NW and SW sub-clusters (see Fig. \ref{fig:hlres}, left). One of these is the brightest cluster galaxy (BCG) belonging to the SE sub-cluster \citep[e.g.][]{Russell2011}. In projection, the radio galaxies are approximately coincident with the extended non-thermal emission from the ICM \citep{Hlavacek-Larrondo2017} and are situated behind the SE and NW merger shocks \citep[e.g.][]{Russell2011}. The total flux density of these sources at 144 MHz is $75\pm11\,\text{mJy}$ for A2146-A (i.e. the SE BCG) and $166\pm25\,\text{mJy}$ for A2146-B. We also obtained measurements at 150 MHz from the TGSS-ADR1 \citep[][]{Intema2017}, 325 MHz \citep{Russell2011}, 612 MHz (see Sec. \ref{sec:red_gmrt}), VLA 1.5 GHz \citep{Hlavacek-Larrondo2017}, and 5 GHz \citep{Hogan2015,Hlavacek-Larrondo2017} observations and present the spectral properties of these sources in Fig. \ref{fig:int_ps} and Table \ref{tab:int_flux}. Since both A2146-A and A2146-B are present in the TGSS-ADR1 150 MHz image, but A2146-A resides in the region that has high background and is only detected at $\lesssim2\sigma$, we exclude the TGSS-ADR1 flux density measurements for A2146-A in this analysis. 

%

\begin{table}
        \centering      
        \caption{Flux density for the radio galaxies in the vicinity of A2146.}
        \label{tab:int_flux}
        \begin{tabular}{lcccc}
                \hline\hline
                Freq. & $S_\text{A2146-A}$ & $S_\text{A2146-B}$ & Telescope &    Ref.    \\
                (MHz) &       (mJy)        &       (mJy)        &           &  \\ \hline
                144   &     $75\pm11$      &     $166\pm25$     &   LOFAR   & this paper \\
                150   &        $-$         &     $183\pm21$     &   GMRT    &    $a$     \\
                238   &    $56.8\pm5.7$    &   $106.1\pm10.6$   &   GMRT    & this paper \\
                325   &      $47\pm5$      &      $93\pm9$      &   GMRT    &    $b$     \\
                612   &    $37.8\pm3.8$    &    $67.6\pm6.8$    &   GMRT    & this paper \\
                1400  &    $15.6\pm3.5$    &    $36.6\pm4.0$    &    VLA    &    $a$     \\
                5000  &    $6.6\pm0.3$     &    $8.9\pm0.5$     &    VLA    &   $c,a$    \\ \hline\hline
        \end{tabular} \\
        Notes: $a$: \cite{Hlavacek-Larrondo2017}, $b$: \cite{Russell2011}, $c$: \cite{Hogan2015}
\end{table}

To estimate the integrated spectral indices for A2146-A and A2146-B, we fit their spectra with an exponential function, $S\propto\nu^\alpha$. In the fitting, the flux densities are weighted by the inverse square of the flux density errors. We find that the average spectral index is $-0.71\pm0.02$ for A2146-A and $-0.84\pm0.02$ for A2146-B. However, the spectral energy distribution in Fig. \ref{fig:int_ps} hints at possible spectral breaks at about 612 MHz and 1.5 GHz for A2146-A and A2146-B, respectively. To quantify this, we fit the radio flux density of the galaxies with a double power-law function. The resulting spectral indices are $\alpha_\text{144\,MHz}^\text{612\,MHz}=-0.53\pm0.11$ and $\alpha_\text{612\,MHz}^\text{5\,GHz}=-0.79\pm0.05$ for A2146-A and $\alpha_\text{144\,MHz}^\text{1.5\,GHz}=-0.64\pm0.06$ and $\alpha_\text{1.5\,GHz}^\text{5\,GHz}=-1.17\pm0.10$ for A2146-B, where in both cases the spectral index steepens at  high frequencies. The spectral steepening of A2146-A and A2146-B  at high frequencies might be due to the synchrotron and IC energy losses of the radio emitting particles.

It is noted that the spectrum below 612 MHz for A2146-A roughly follows a single power-law function, $S\propto\nu^\alpha$, and does not indicate a sharp turnover at low frequencies as reported in \cite{Hlavacek-Larrondo2017}, which used GMRT 150 MHz data.

\subsection{North-west extended emission}
\label{sec:res_dis_relics}

In Fig. \ref{fig:hlres}, the NW radio emission extends over a region of $200\times310\,\text{kpc}^2$. The NW radio emission is elongated in the NE-SW direction, which is approximately perpendicular to the merger axis \citep[e.g.][]{White2015}. The outer edge of the NW radio emission is located behind the detected upstream X-ray shock \citep{Russell2010}. The integrated flux densities of the NW emission measured within the $2.5\sigma$ contours are $S_\text{144\, MHz}=13.1\pm2.0\,\text{mJy}$ and $S_\text{1.5\,GHz}=0.89\pm0.08\,\text{mJy}$, resulting in an integrated spectral index of $\alpha=-1.14\pm0.08$. This spectral index measurement for the NW radio emission is typical value for a known elongated relic \citep[i.e. $\overline{\alpha}\approx-1.3$;][]{Luigina2012}. However, our spectral index is flatter than the 1-2 GHz in-band measurement of $\alpha_\text{1-2\,GHz}=-2.3\pm0.3$ that was reported in \cite{Hlavacek-Larrondo2017}. As mentioned by \cite{Hlavacek-Larrondo2017}, the true uncertainty of the in-band spectral measurement is likely higher. However, if the true value of the 1-2 GHz spectral index of the NW emission is indeed this steep then it would imply a spectral curvature between 144 MHz and 1.5 GHz. Unfortunately, we are unable to check this possibility because the NW extended emission has only been detected at 144 MHz and 1.5 GHz.

\begin{figure*}
        \centering
        \includegraphics[width=0.5\textwidth]{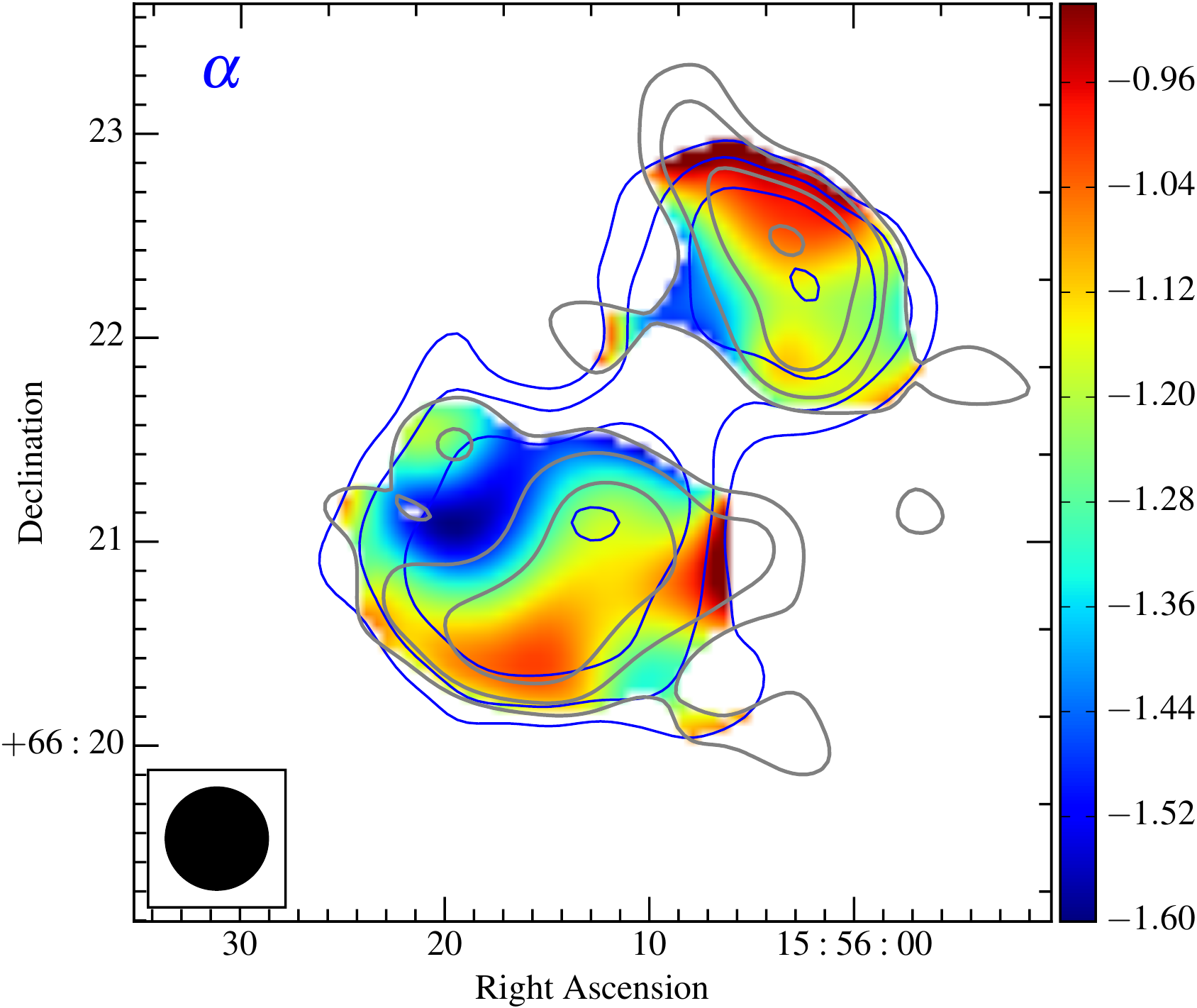} \hfill
        \includegraphics[width=0.49\textwidth]{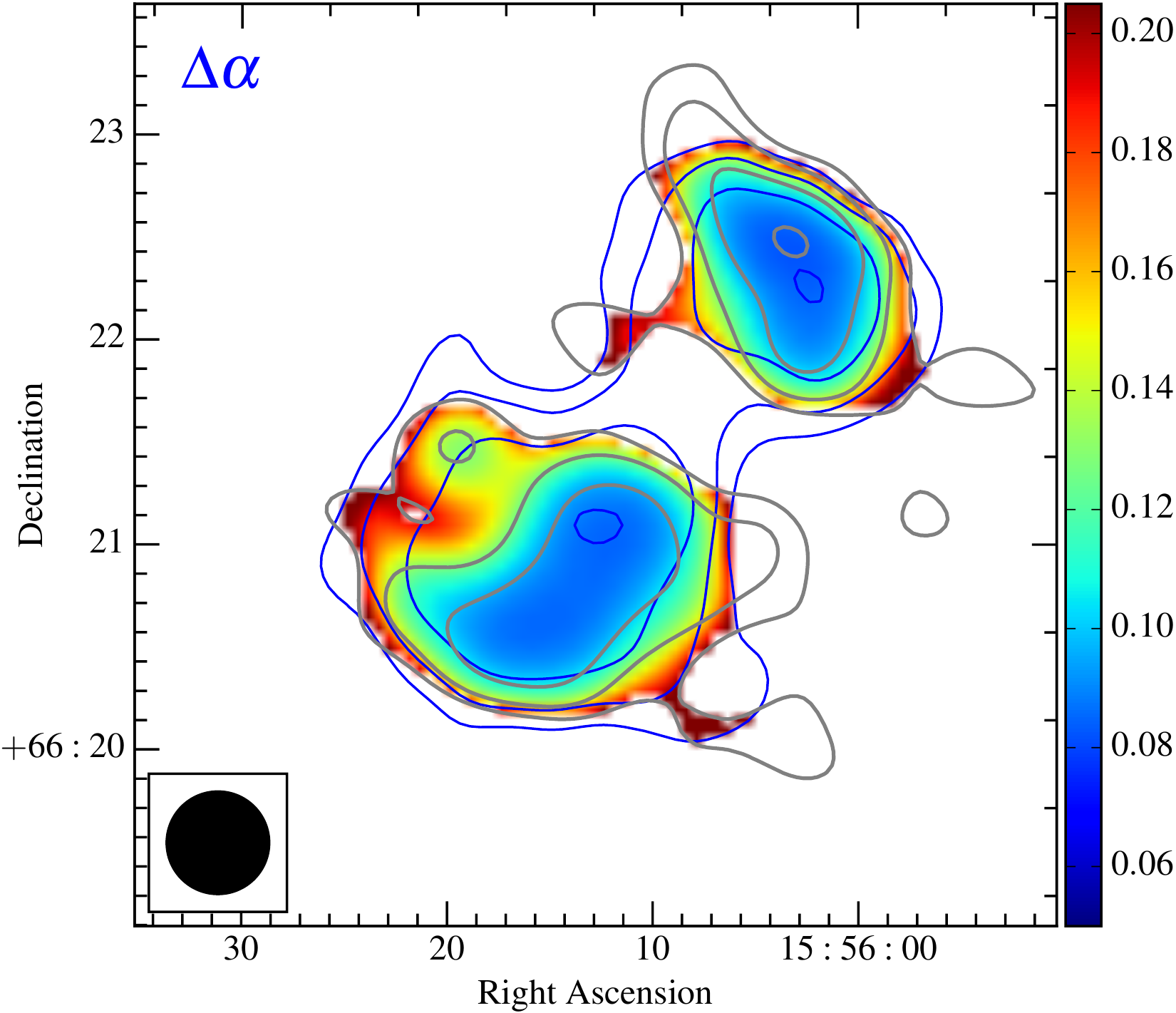}
        \caption{Left: Spectral index map between $144\,\text{MHz} - 1.5\,\text{GHz}$ for the extended radio emission from A2146. Right: the corresponding spectral index error map. The resolution of the maps is $30\arcsec$. In both images, the LOFAR 144 MHz (blue) and VLA 1.5 GHz (grey) first contours are $2.5\sigma$, where $\sigma=340$ and $27\,\upmu\text{Jy beam}^{-1}$ for the LOFAR 144 MHz and VLA 1.5 GHz data, respectively. The next contours are multiplied by $\sqrt{3}$.}
        \label{fig:spx}
\end{figure*}

\begin{figure}
        \centering
        \begin{minipage}{0.48\textwidth}
                \centering
                \begin{tikzpicture}
                        \node[anchor=south west,inner sep=0] (image) at (0,0) {\includegraphics[width=1\columnwidth]{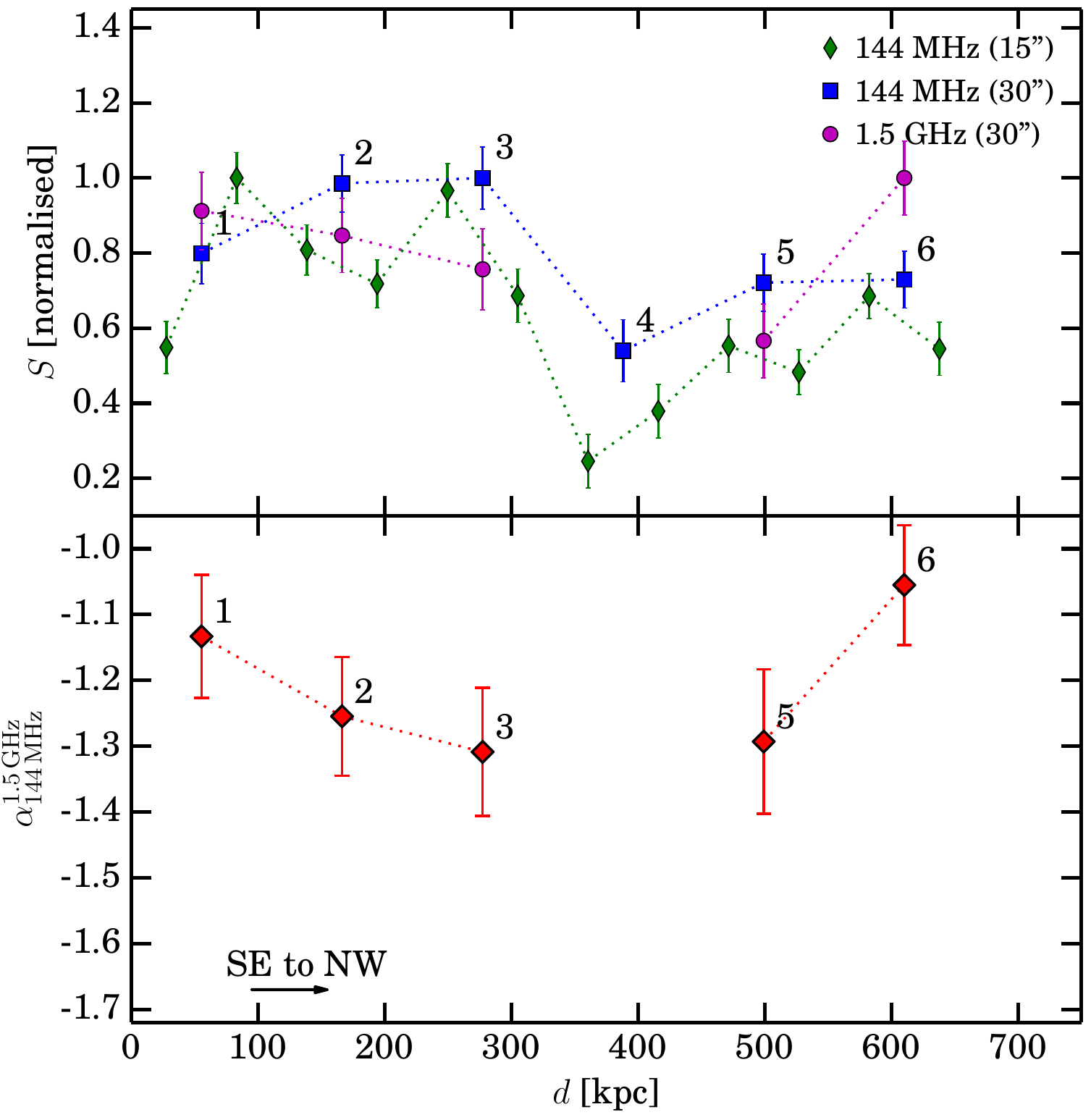}};
                        \begin{scope}[x={(image.south east)},y={(image.north west)}]
                        \node[anchor=south west, inner sep=0] (image) at (0.73,0.1) {\includegraphics[width=0.25\columnwidth]{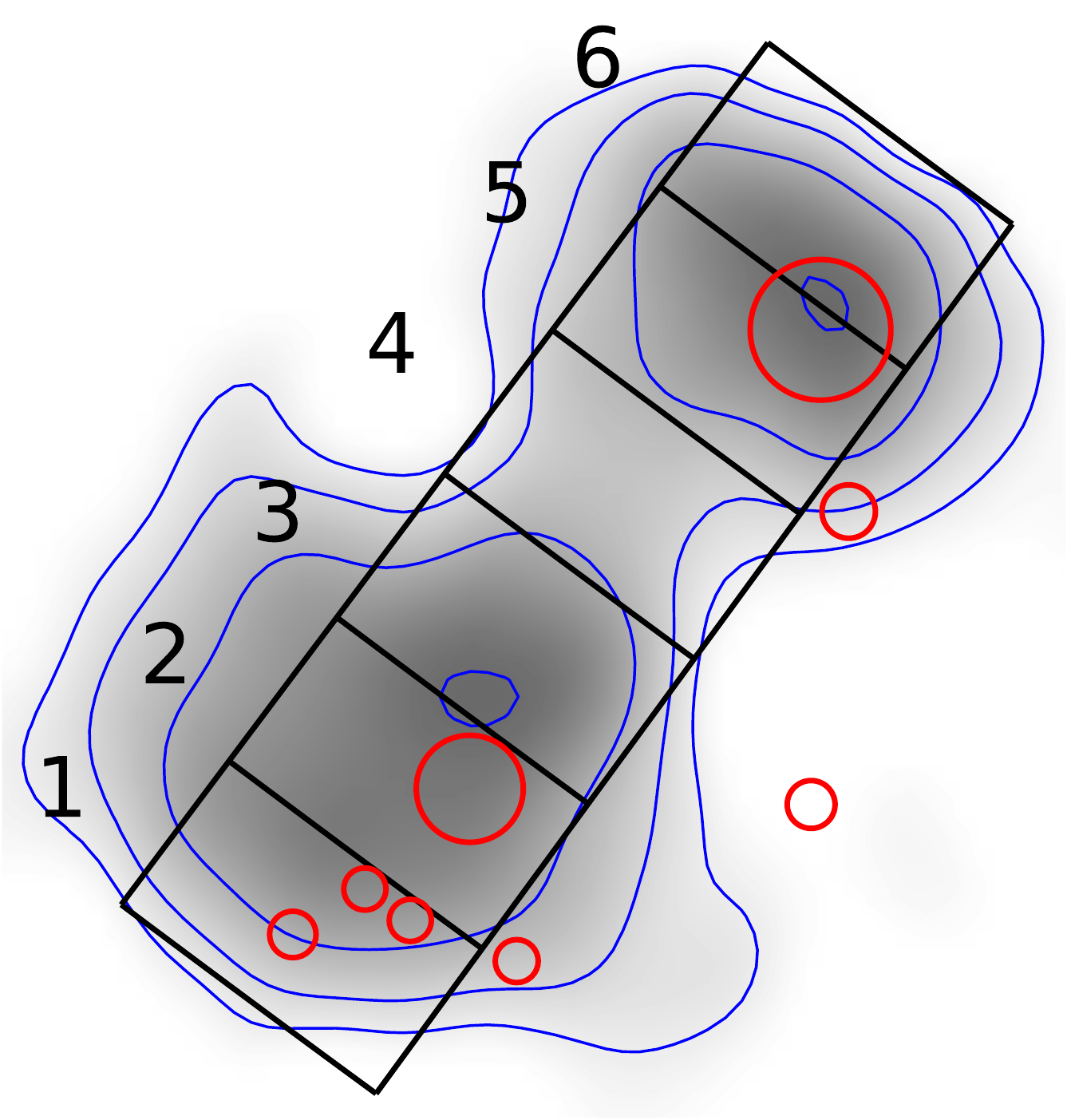}};
                        \end{scope}
                \end{tikzpicture}
        \end{minipage}
        \caption{Surface brightness (top) and spectral index (bottom) profiles for the extended radio emission from A2146. The flux scale errors are not added to the total errors of the flux densities. The numbers next to the data points indicate the region numbers shown in overlaid image. The red circular regions at the locations of the subtracted compact sources are masked.}
        \label{fig:sb_spx_pro}
\end{figure}

To examine the distribution of the spectral index and search for a spectral index gradient that is often associated with radio relics, we make the spectral index map in Fig. \ref{fig:spx} and the spectral index profile in Fig. \ref{fig:sb_spx_pro}. In the spectral index profile a  steepening of the spectral index of the NW emission  approximately along the merger axis is apparent. In the outer region, the spectral index is $-1.06\pm0.06$, but steepens to $-1.29\pm0.09$ in the inner region about $170\,\text{kpc}$ from the outer $2.5\sigma$ contour. Since we examine the spectral trend, the flux scale errors due to the amplitude calibration are not used in the calculation of the spectral index errors. We emphasise that because of the limited resolution of our spectral index map we are only able to make two independent measurements of the spectral index in the region of NW emission and that it is important to confirm the spectral gradient across the NW extended radio emission with high-resolution observations.

Basing on the morphology of the NW radio emission, \cite{Hlavacek-Larrondo2017} concluded that it is likely a radio relic. A number of radio relics have been observed to have steeper spectra towards the cluster centre \citep[e.g.][]{Bonafede2009,Kale2012b,DeGasperin2015,VanWeeren2010a,VanWeeren2017}. The spectral steepening is due to the energy losses through the synchrotron and IC emission. In the spectral analysis above, we show that the radio spectral index in the region behind the NW shock front is steeper than that in the outer edge, which supports the nature of the NW extended radio emission being a radio relic. Its location and spectral properties imply that the NW extended emission is likely to be associated with the NW upstream shock. Shocks compress ICM magnetic fields resulting in polarised radio emission. Therefore, new radio polarisation observations will provide further information concerning the relation between the shock and extended emission in the NW region of the cluster.

According to the diffusive shock acceleration (DSA) \citep[e.g.][]{Blandford1987}, a shock of Mach number $\mathcal{M}$ should accelerate thermal electrons and generate a population of relativistic electrons with an energy spectrum, 
\begin{equation}
         \frac{dN}{d\gamma}=N_0\gamma^{-\delta_\text{inj}},
         \label{eq:spec}
\end{equation}
where $N_0$ and $\gamma$ are the spectral normalisation and Lorentz factors, respectively; the spectral slope $\delta_\text{inj}$ is related to the shock Mach number via
\begin{equation}
        \delta_\text{inj}=2\frac{\mathcal{M}^2+1}{\mathcal{M}^2-1}.
        \label{eq:delta}
\end{equation}
In the presence of magnetic fields, the relativistic electrons emit synchrotron radiation with a spectrum of $I\propto\nu^{\alpha_\text{inj}}$, where $\alpha_\text{inj}=(1-\delta_\text{inj})/2$. The NW upstream shock in A2146 has a Mach number of $\mathcal{M}=1.6\pm0.1$ that was derived from the X-ray SB jump \citep[][]{Russell2012}. According to the DSA model, we should therefore observe radio emission along the shock with an injection spectral index of $\alpha_\text{inj}=-1.78^{+0.22}_{-0.32}$. This prediction is steeper than our estimate, i.e. $\alpha_\text{inj}=-1.06\pm0.09,$ which takes into account the flux scale errors of 15\% and 5\% for LOFAR and VLA, respectively. The discrepancy should not be caused by the resolution of the radio observations because if there is indeed a spectral index  gradient at higher resolution the index would be even flatter in the region closer to the shock front. The discrepancy could be because the shock is not powerful enough to accelerate thermal electrons efficiently and is instead \mbox{(re-)accelerating} pre-existing fossil electrons. Such electrons could be remnants of lobes of radio galaxies \cite[e.g.][]{Markevitch2005,Kang2012}. In this scenario, the shock increases the brightness of the radio emission, but preserves the spectrum of the pre-shock fossil electrons if the fossil electrons have a flatter spectrum than that generated by the shock \citep[e.g.][]{Markevitch2005}. If fossil electrons were being \mbox{re-accelerated} we may see extended radio emission in front of the shock \citep{Markevitch2005} but this is not present in the LOFAR 144 MHz and VLA 1.5 GHz observations (Figs. \ref{fig:hlres} and \ref{fig:spx}). Finally, we note that the spectral index of $-1.06$ that we estimated for the NW outer region is the typical value for the lobes of radio galaxies, although unlike \cite{VanWeeren2017} the potential source of the fossil electrons in the NW region of the cluster is not obvious. 


As discussed in \cite{VanWeeren2016b} for the Toothbrush
relic, an alternative explanation for the mismatch of injection indices obtained from the radio observations and the DSA model is that the shock might contain different Mach numbers along the line of sight \citep[e.g.][]{Skillman2011,Skillman2013,Vazza2012}. In the case of a nonlinear dependence of the Mach numbers on the acceleration efficiency \citep[e.g.][]{Hoeft2007}, the radio observations are particularly sensitive to the shocks with high Mach numbers (resulting in flat spectrum radio emission), while the X-ray observations tend to observe the lower Mach number shocks (corresponding to steep spectrum radio emission). This explanation is supported by recent simulations of particle acceleration at cluster merger shocks in \cite{Hong2015} and \cite{Ha2017a}. These studies reported that the weighted Mach numbers derived from the gas temperature jumps are smaller than those obtained from the shock kinetic energy flux (i.e. radio data). Hence, the discrepancy between the injection indices derived from spectral index map and the DSA model might be explained if the shock consists of multiple Mach numbers along the line of sight. We note that if we use the Mach number $\mathcal{M}_\text{NW}=2.0\pm0.3$ calculated from temperature jump \citep{Russell2012}, the injection index predicted by the DSA model is $\alpha=-1.17^{+0.20}_{-0.39}$, which is consistent with our injection spectral measurement (i.e. $\alpha_\text{inj}=-1.06\pm0.09$). However, \cite{Russell2012} pointed out that the Mach number calculated from the temperature jump is less accurate owing to the large radial bins that cannot resolve the shock jump accurately. 

The Mach numbers calculated from X-ray SB density jump could be biased low because of projection effects. Without the correction for the projection effects, this bias increases the difference between the X-ray and radio derived Mach numbers (or the DSA predicted and observed injection indices). However, as two sub-clusters in A2146 are known to merge almost on the plane of the sky (i.e. the angle between the merger axis and the plane of the sky is $\sim16^\circ$;  \citealt{White2015}), the viewing angle to the NW shock is likely to be edge on. Hence, the projection effects might not play an important role in the mismatching of the DSA predicted and radio observed injection indices in the NW extended emission. 


\begin{figure}
        \centering
        \includegraphics[width=0.48\textwidth]{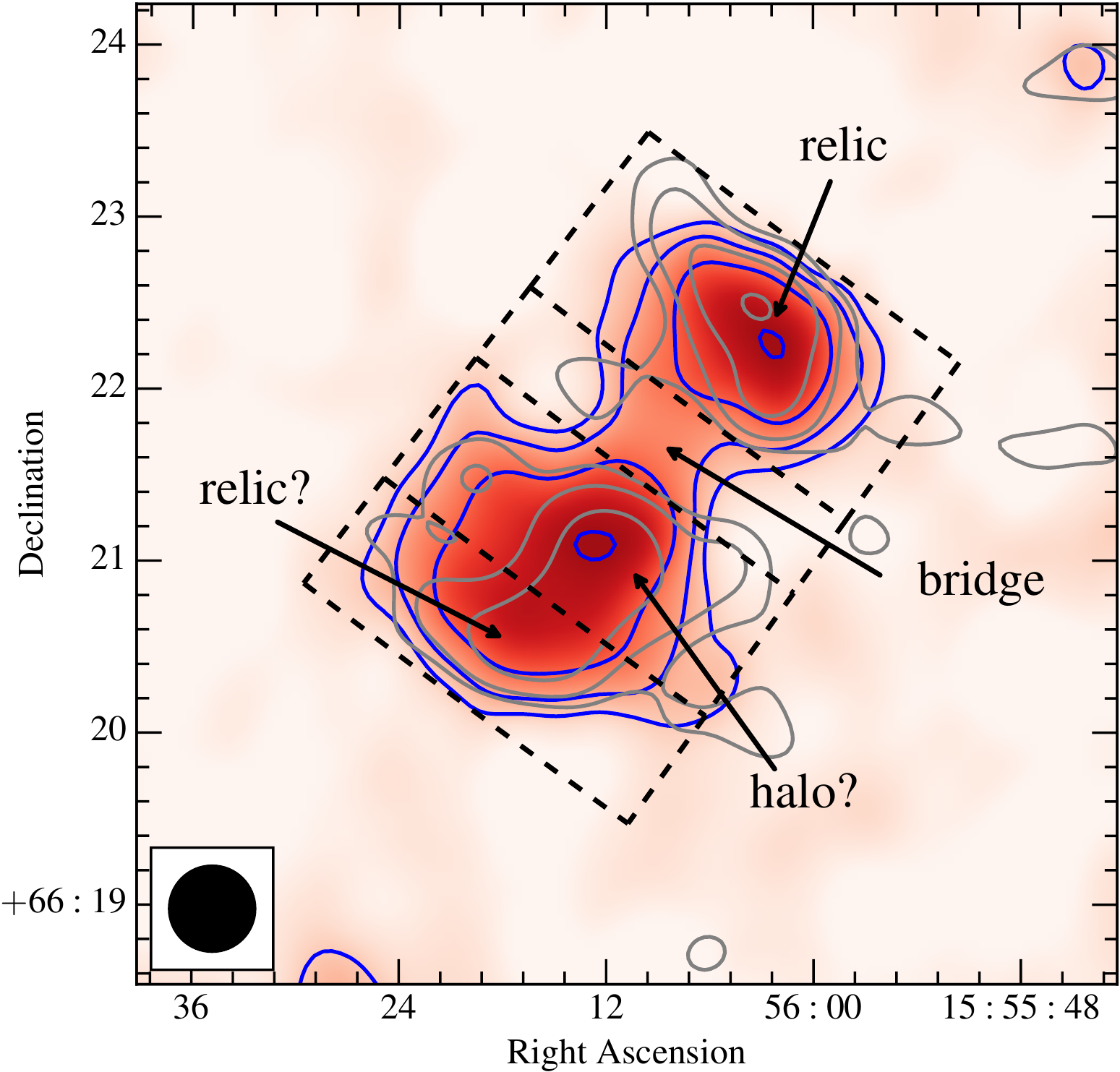}
        \caption{Regions (dashed lines) in which flux densities were extracted are shown on the LOFAR 144 MHz $30\arcsec$resolution image. The LOFAR 144 MHz (blue) and VLA 1.5 GHz (grey) contours are identical to those in Fig. \ref{fig:spx}. }
        \label{fig:sources}
\end{figure}

\subsection{Radio bridge}
\label{sec:bridge}

In Fig. \ref{fig:hlres}, the NW relic is connected with the SE extended emission through a faint emission (bridge). We measure the flux density of the radio emission in the bridge using the $>2.5\sigma$ pixels within the region of the bridge (see Fig. \ref{fig:sources}). The integrated flux density of the emission in the bridge is $1.1\,\text{mJy}$ at 144 MHz, but it is undetected at 1.5 GHz (see Figs. \ref{fig:spx} and \ref{fig:sources}). We assume the size of the bridge emission at 1.5 GHz is similar to that at 144 MHz. We find that $1\sigma$ upper limit for the bridge emission at 1.5 GHz is $0.047\,\text{mJy}$. This implies that the integrated spectral index of the emission in the bridge is likely steeper than $-1.3$. The elongated morphology of the bridge along the merger axis implies that it likely associates with the turbulence induced by the ongoing merger. However, it is still unclear whether or not the bridge emission is a separated source or is part of the SE extended emission.

\subsection{South-east extended emission}
\label{sec:res_dis_halo}

In Fig. \ref{fig:hlres}, we show the SE extended emission at 144 MHz. The SE emission has a projected size of $300\times350\,\text{kpc}^2$ and is elongated in the NW-SE direction. The SE edge of the SE emission roughly follows the SE bow shock \cite[e.g.][]{Russell2010}. The integrated flux density of the SE extended emission (without the bridge, see Figs. \ref{fig:hlres} and \ref{fig:sources}) encompassing $2.5\sigma$ contours is $24.3\pm3.8\,\text{mJy}$ at 144 MHz and $1.3\pm0.1\,\text{mJy}$ at 1.5 GHz. The spectral index between 144 MHz and 1.5 GHz is $\alpha=-1.25\pm0.07$. Unlike the spectral index estimate for the NW relic, our spectral index measurement for the SE emission is consistent with the VLA in-band estimate of $-1.2\pm0.1$ that was presented in \cite{Hlavacek-Larrondo2017}. The spectral index map in Fig. \ref{fig:spx} shows a patchy distribution in the SE region. Along the SE-NW merger axis, the SE extended emission has a flat spectrum in the SE region and the spectrum becomes steeper towards the NW direction. The trend is more visible in the spectral index profile in Fig. \ref{fig:sb_spx_pro}. 

The nature of the SE extended emission is still unclear. Possibilities include (i) a radio halo bounded by a merger shock \citep[e.g.][]{Markevitch2005,Shimwell2014,Markevitch2010,Brown2011b,VanWeeren2016b} or (ii) a radio relic on the SE edge superimposed on a radio halo that extends outwards from the cluster centre \cite[e.g.][]{Brunetti2008,Macario2011a,VanWeeren2016b,Hoang2017a}. Scenario (ii) was the favoured scenario in \cite{Hlavacek-Larrondo2017}. 
The extended emission in scenarios (i) and (ii) can perhaps be generated by the same merger; but the distinct appearance of the radio emission and its spectrum depends on the shock Mach number, the magnetic field strength behind the shock front, the spectral energy distribution of turbulence after the shock passage, and the observing frequencies \citep{Markevitch2010}. It is noted that the SE edge of the SE extended emission is not spatially coincident with the entire shock front (see Fig. \ref{fig:hlres}, right), although the location and orientation of the SE extended emission along the merger axis seems to imply its shock-related origin.

\begin{figure}
        \centering
        \includegraphics[width=0.48\textwidth]{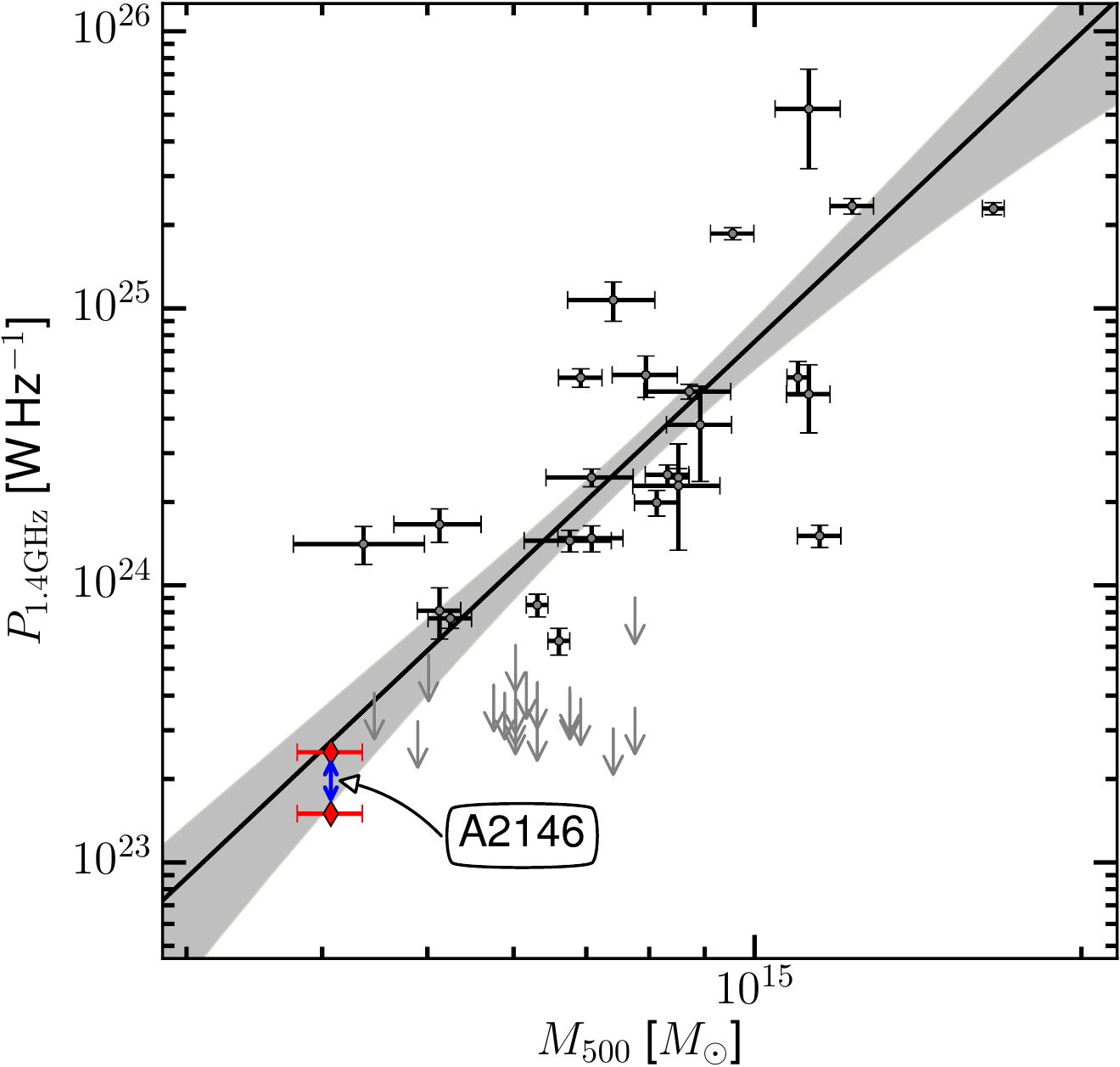}
        \caption{Correlation between $P$ and $M$. The estimated range for the power of the radio halo in A2146 (red points) is roughly consistent with the $P-M$ scaling relation (i.e. black line). The grey shadow is the 95\% confidence region. The downward arrows show upper limits for undetected haloes. For the list of the haloes in the plot, we refer to \cite{Cassano2013a}.}
        \label{fig:PM}
\end{figure}

According to scenario (ii) the SE extended emission (Fig \ref{fig:hlres}) should consist of two components. To assess this we make SB profiles along the merger axis in Fig. \ref{fig:sb_spx_pro}. To take into account the size of the synthesis beam, each data point is estimated in a region that encompasses a beam ($15\arcsec$ or $30\arcsec$). The brightness peaks in the SE extended radio emission are clearest in the $15\arcsec$ resolution profile as in the $30\arcsec$ resolution profile they are smoothed out. We note that the extended emission maps are obtained by subtracting compact sources (i.e. cyan circles in Fig. \ref{fig:hlres}, right), which leave some residual flux that contaminates the profiles. However, if there are two separate sources (i.e. halo and relic) in the SE region, the width of the relic can be approximated as the distance from the SE $2.5\sigma$ contour to the local minimum brightness in the $15\arcsec$ profile, which is about $170\,\text{kpc}$ (Fig. \ref{fig:sb_spx_pro}). With this approximation we are able to estimate the halo flux; these  estimates should be considered as the lower limits for the halo flux density. To have consistent flux measurements at both frequencies, we measure the halo flux densities at 144 MHz and 1.5 GHz using the $30\arcsec$ images within the regions shown in Fig. \ref{fig:sources}. Only pixels that are detected at $>2.5\sigma$, where $\sigma=340$ and $27\,\upmu\text{Jy\,beam}^{-1}$ for the LOFAR 144 MHz and VLA 1.5 GHz images, respectively, are used in the calculation. The integrated flux densities at 144 MHz for the relic and halo regions are $9.0$ and $15.3\,\text{mJy}$, respectively. At 1.5 GHz, they are $0.53$ and $0.78\,\text{mJy}$, respectively. With these measurements, we find the integrated spectral index to be $-1.2$ for the tentative relic and $-1.3$ for the halo. Since the separation of the relic and halo require verification with high-resolution observations, the flux density estimate in this case should be consider as a rough estimate. The true value for the flux density of the halo at 144 MHz should lay in between $15.3\,\text{mJy}$ (i.e. the SE extended emission consists of a relic and a halo) and $24.3\,\text{mJy}$ (i.e. the SE extended emission is a single halo). At 1.5 GHz, it is between $0.78\,\text{mJy}$ and $1.31\,\text{mJy}$. If the bridge emission is part of the SE extended emission, the upper limit for the flux density of the halo is $25.4\,\text{mJy}$ at 144 MHz and $1.38\,\text{mJy}$ at 1.5 GHz. The 1.4 GHz power for radio haloes, which is proportional to the amount of turbulence energy that is converted into the relativistic electrons, is known to correlate with the masses of their host clusters \citep[e.g.][]{Cassano2013a}. Using the flux density estimates above for the radio halo, we calculate the power of the A2146 halo at 1.4 GHz is in the range $(1.5-2.5)\times10^{23}\,\text{W\,Hz}^{-1}$ (\textit{k}-corrected). The power of the A2146 halo is consistent with the $P-M$ scaling relation with a cluster mass of $M_\text{500}=(4.04\pm0.27)\times10^{14}\,\text{M}_\odot$ \citep{Planck2015} that we present in Fig. \ref{fig:PM}. 

Another key observational feature to distinguish whether the extended emission is a single halo (i.e. the case i) or a composition of a halo and a relic (i.e. the case ii) is the distribution of the spectral index of the radio emission along the axis of shock propagation. In case (ii), we expect the spectral index of the radio emission is flat immediately behind the shock front (i.e. due to shock \mbox{(re-)acceleration} or compression), steepening  further in the post-shock region (i.e. because of synchrotron and IC losses) and might flatten again in the halo region depending on the amount of turbulent energy induced by the passage of a merger shock available in the ICM. On the other hand, in case (i) whether or not we might see a similar spectral trend across the extended source behind the shock, depending on the physical conditions at the post-shock region. The spectral index of the halo emission might slightly steepen behind the shock and then become uniform across the halo region. These scenarios (i) and (ii) have recently been used to explain the spatial distribution of the spectral index in some clusters. Examples for case (i) are the SW radio edge in Abell 520 \citep[][submitted]{Hoang2018b} and, perhaps, the southern region of the radio halo in the Toothbrush cluster \citep{VanWeeren2016b,Rajpurohit2018}. For case (ii), examples are the northern relics and the haloes in the Toothbrush and Sausage clusters \citep[e.g.][]{VanWeeren2016b,Hoang2017a}. 
Whilst our spectral index maps are too low resolution to make firm conclusions, in the SE extended emission of A2146, there might be a spectral gradient from $-1.13\pm0.06$ to $-1.25\pm0.06$ behind the shock. This tentative trend is in line with that expected for a  relic-halo superposition. If the SE extended radio emission is shock \mbox{(re-)accelerated}, the radio spectral index of $-1.13\pm0.06$  in the SE edge implies a shock Mach number of $\mathcal{M}_\text{SE}=2.0\pm0.1$. This is in line the value estimated from the X-ray data \citep[i.e. $\mathcal{M}_\text{SE}^\text{X}=2.3\pm0.2$;][]{Russell2012}, which makes the argument for a connection between the bow shock and SE edge of the extended radio emission (i.e. the relic) in the SE region of the cluster more compelling.

\section{Conclusions}
\label{sec:conclusions}

In this paper, we present the results of deep LOFAR $120-168$ MHz observations of the binary merging galaxy cluster A2146. We map the extended continuum emission at 144 MHz and the spectral energy distribution in the cluster in more detail than in previous studies. We summarise the results below.

\begin{itemize}

%
        \item[$\bullet$] The LOFAR 144 MHz observations confirm the presence of the NW extended emission that was detected with the deep VLA 1.5 GHz observations \citep{Hlavacek-Larrondo2017}. The radio emission extends behind the upstream shock front and have a flux density of $13.1\pm2.0\,\text{mJy}$ at 144 MHz and $0.89\pm0.08\,\text{mJy}$ at 1.5 GHz. The integrated spectral index is $\alpha=-1.14\pm0.08$. The spectral index flattens to $-1.06\pm0.06$ in the outer region and steepens to $-1.29\pm0.09$ in the inner region. The morphological and spectral properties of the NW extended emission are consistent with the hypothesis that the NW extended emission is a relic associated with the NW upstream merger shock.
        \item[$\bullet$] The DSA model predicts an injection spectral index of $-1.78^{+0.22}_{-0.32}$ for the $\mathcal{M}_\text{NW}=1.6\pm0.1$ NW shock. However, we measure a spectral index in the outer region of the NW relic to be $-1.06\pm0.09$ (taken into account flux scale errors), which is inconsistent with the DSA prediction. The mismatching of the injection spectrum indices could be explained if the shock \mbox{re-accelerates} a pre-existing population of fossil electrons rather than those in the thermal pool.
        \item[$\bullet$] With the LOFAR 144 MHz observations, we detect a faint emission (bridge) connecting the NW and SE regions. The integrated flux density of the radio emission in the bridge is $1.1\,\text{mJy}$ at 144 MHz. The non-detection of the bridge with the VLA 1.5 GHz observations implies that the spectral index of the bridge emission is steeper than $-1.3$.
        \item[$\bullet$] The SE extended emission has an integrated flux density of $24.3\pm3.8\,\text{mJy}$ at 144 MHz and $1.3\pm0.1\,\text{mJy}$ at 1.5 GHz, resulting in $\alpha=-1.25\pm0.07$. Further analysis of the brightness suggests that the SE extended emission may consist of a halo in the central region and a relic in the SE edge. The power for the radio halo is constrained within  $(1.5-2.5)\times10^{23}\,\text{W\,Hz}^{-1}$, which is roughly consistent with the expected power for the cluster mass \citep[i.e.  $M_\text{500}=(4.04\pm0.27)\times10^{14}\,\text{M}_\odot$;][]{Planck2015}, according to the $P-M$ scaling relation \citep{Cassano2013a}.

%
        
\end{itemize}

\section*{Acknowledgments}

DNH and HR acknowledge support from the ERC Advanced Investigator programme NewClusters 321271.
RJvW acknowledges support from the VIDI research programme with project number 639.042.729, which is financed by the Netherlands Organisation for Scientific Research (NWO). The LOFAR group in Leiden is supported by the ERC Advanced Investigator programme New-Clusters 321271. 
This paper is based (in part) on data obtained with the International LOFAR
Telescope (ILT) under project code LC7\_024 and DDT9\_001. The LOFAR instrument (van Haarlem et al. 2013) is the Low
Frequency Array designed and constructed by ASTRON. This instrument has observing, data
processing, and data storage facilities in several countries; these facilities are owned by
various parties (each with their own funding sources) and are collectively
operated by the ILT foundation under a joint scientific policy. The ILT resources
have benefitted from the following recent major funding sources: CNRS-INSU,
Observatoire de Paris and Université d'Orléans, France; BMBF, MIWF-NRW, MPG,
Germany; Science Foundation Ireland (SFI), Department of Business, Enterprise and
Innovation (DBEI), Ireland; NWO, The Netherlands; The Science and Technology
Facilities Council, UK; and Ministry of Science and Higher Education, Poland. 
Part of this work was carried out on the Dutch national e-infrastructure with the support of the SURF Cooperative through grant e-infra 160022 and 160152.  The LOFAR software and dedicated reduction packages on https://github.com/apmechev/GRID\_LRT were deployed on the e-infrastructure by the LOFAR e-infragroup, consisting of J. B. R. Oonk (ASTRON \& Leiden Observatory), A. P. Mechev (Leiden Observatory) and T. Shimwell (ASTRON) with support from N. Danezi (SURFsara) and C. Schrijvers (SURFsara).
We thank the staff of the GMRT who made these observations possible. The GMRT is run by the National Centre for Radio Astrophysics of the Tata Institute of Fundamental Research.
The National Radio Astronomy Observatory is a facility of the National Science Foundation operated under cooperative agreement by Associated Universities, Inc. 
The scientific results reported in this article are based in part on data obtained from the Chandra Data Archive and observations made by the Chandra X-ray Observatory and published previously in cited articles. 
%
We thank Helen Russell for providing the Chandra X-ray image of Abell 2146 and commenting on the manuscript.



\bibliographystyle{aa}
\bibliography{library}

%

\end{document}